\documentclass[aps,showpacs,
twocolumn,
preprintnumbers,
nofootinbib,
citeautoscript,
prl]{revtex4}

\IfFileExists{srcltx.sty}{\usepackage[active]{srcltx}}%
\usepackage{epsfig}
\usepackage{graphicx}
\usepackage{epsfig}
\usepackage{bm}
\usepackage{amsfonts,amssymb,amsmath} 
\usepackage{xspace}
\usepackage[first,bottomafter,timestamp]{draftcopy}
 
\newcommand{\beq}{\begin{equation}}
\newcommand{\eeq}{\end{equation}}
\newcommand{\bea}{\begin{eqnarray}}
\newcommand{\eea}{\end{eqnarray}}

\renewcommand{\ol}{\overline}
\def\simle{\lower 2pt \hbox {$\buildrel < \over {\scriptstyle \sim }$}}
\def\simge{\lower 2pt \hbox {$\buildrel > \over {\scriptstyle \sim }$}}

\begin{document}

\preprint{TU-764, UCLA/06/TEP/05, CERN-PH-TH/2006-023}

\title{Opening a new window for warm dark matter}

\author{Takehiko Asaka$^1$,  Mikhail
Shaposhnikov$^{2,3}$ and  Alexander Kusenko$^4$}

\affiliation{
$^1$Department of Physics, Tohoku University, Sendai 980-8578, Japan\\
$^2$Institute of Theoretical Physics, \'Ecole Polytechnique
F\'ed\'erale de Lausanne, CH-1015 Lausanne,
Switzerland\\ 
$^3$Theory Division, CERN, CH-1211 Geneva 23, Switzerland\\
$^4$Department of Physics and Astronomy, University of California, Los
Angeles, CA 90095-1547, USA 
}

%\date{January, 2006}

\begin{abstract}

We explore the range of parameters for dark-matter sterile neutrinos
in an extention of the Minimal Standard Model by three singlet
fermions with masses below the electroweak scale (the $\nu$MSM). 
This simple model can explain a wide range of phenomena, including
neutrino oscillations, baryogenesis, the pulsar velocities, and the
early reionization.  The presence of two heavier sterile neutrinos
and the possibility of entropy production in their decays broadens
the allowed range of parameters for the dark-matter sterile neutrinos
(or other types of dark matter, for example, the gravitino). In
addition, the primordial production of dark matter sterile neutrinos
allows to escape most of the constraints.

\end{abstract}


\maketitle

\renewcommand{\thefootnote}{\arabic{footnote}}
\setcounter{footnote}{0}

{\bf Introduction.} While the Minimal Standard Model (MSM) \cite{msm}
of electroweak interactions is in good agreement with accelerator
experiments, it is known to be incomplete in the neutrino sector.
Neutrino oscillations have been observed in a number of experiments
demonstrating that neutrinos have non-zero masses and mixings (for a
review of the current constrains on neutrino mass matrix see, e.g.
\cite{Strumia:2005tc}). In addition, the origin of matter-antimatter
asymmetry and the nature of dark matter have no explanation in the
framework of the MSM. It has been shown recently that all these three
phenomena can be explained in a simple extension of the Standard
Model achieved by adding three gauge singlet fermions, the sterile
neutrinos, with masses smaller than the electroweak scale
\cite{Asaka:2005an,Asaka:2005pn}. One of these sterile neutrinos can
be dark matter, while the other two allow for baryogenesis. This
model was called $\nu$MSM for the addition of extra neutrinos to the
MSM.  

The motivation for the $\nu$MSM is further strengthened by the
preponderance of indirect hints, each of which individually may not
be very compelling, but which collectively make a strong case for
taking the model seriously. As shown in Ref.~\cite{Asaka:2005an}, the
minimal number of sterile neutrinos that can explain dark matter in
the universe, while being consistent with the experimental data on
neutrino oscillations, is ${\cal N}=3$.  The lightest of these three
states ($N_1$) can be produced in the early universe in just the
right amount to account for cosmological dark matter.  The same
particle would be emitted from a supernova explosion with the
anisotropy that is sufficient to explain the observed velocities of
pulsars~\cite{ks97,fkmp}. X-ray photons from the decays of these
particles can speed up the formation of molecular hydrogen and
precipitate the early star formation and
reionization~\cite{Biermann:2006bu}, in agreement with the WMAP
observations~\cite{WMAP-polar}. The two heavier sterile neutrinos
($N_2$ and $N_3$) can lead \cite{Asaka:2005pn} to generation of
asymmetries between sterile neutrinos and left-handed leptons via
sterile neutrino oscillations~\cite{Akhmedov:1998qx}.    The
asymmetry in left-handed leptons is then converted to baryon
asymmetry due to the anomalous electroweak fermion number
nonconservation at high temperatures~\cite{Kuzmin:1985mm}. 

Given that several seemingly unrelated phenomena are explained in the
framework of a simple model, it is worthwhile to examine the full
range of consequences of  $\nu$MSM. It has been already shown that an
absolute scale of active neutrino masses in this model can be fixed
\cite{Asaka:2005an,Boyarsky:2006jm}. In addition, a constraint on the
rate of neutrinoless double $\beta$ decay has been derived
\cite{Bezrukov:2005mx}.

In this paper we will show that that the range of parameters for warm
dark matter (WDM) in the $\nu$MSM is wider that was previously
determined \cite{Olive:1981ak,Dodelson:1993je,Dolgov:2000ew,
Abazajian:2001nj,Abazajian:2001vt,Abazajian:2002yz,
Hansen:2001zv,Viel:2005qj,Abazajian:2005xn,Abazajian:2005gj} for
dark-matter sterile neutrinos in the context of an extention of the
MSM by just {\it one} sterile neutrino.  In particular, we will
demonstrate that the dark-matter neutrino is allowed to have a
smaller mass  and a larger mixing angle with active neutrinos in
$\nu$MSM.   This has important implications for experimental
detection of dark matter. Moreover, we will argue that the dark
matter sterile neutrino momentum distribution function can be more
complicated than it was previously thought; this result has
ramifications for structure formation in WDM cosmology. In addition,
we will show that the constraints on other WDM particle candidates
(such as light gravitino), which were in thermal equilibrium above
the electroweak scale, are relaxed. We also point out that if the
primordial production of sterile neutrinos takes place, no upper
limit on their mass can be set.

{\bf WDM in the model with one sterile neutrino.}  Let us summarize
here the constraints on mass $M_s$ and mixing angle $\theta$ of the
WDM particle derived for the Standard Model augmented by only one
sterile neutrino\footnote{Note that this model cannot accommodate the
data on neutrino oscillations since two active neutrinos are
predicted to be massless.}. This particle will be denoted by $N_1$.  

First, the mixing angle $\theta$ between an active and a sterile
neutrino must be small enough to avoid overclosing the universe by
sterile neutrinos produced in active-sterile neutrino
oscillations~\cite{Dodelson:1993je,Dolgov:2000ew,Abazajian:2001nj}.
According to \cite{Abazajian:2005gj}, the corresponding limit can be
presented in the form:
\beq
\theta < \theta_{max}(M_s)=
1.3 \times 10^{-4} \left(\frac{1~\mbox{keV}}{M_s}\right)^{0.8}
\label{angle}
\eeq
for the cental value of the parameters for the dark matter abundance 
$\Omega_{DM}=0.22$ and the QCD cross-over transition temperature
$T_{QCD}=170$ MeV. For such small mixing angles, the sterile
neutrinos are out of thermal equilibrium for the entire range of
temperatures at which the MSM with one extra fermion can remain a
good effective theory. Indeed, the maximum of sterile neutrino
production occurs around the temperature \cite{Dodelson:1993je}
\beq
T_p \simeq
130 \left(\frac{M_s}{1~\mbox{keV}}\right)^{\frac{1}{3}}~\mbox{MeV},~
\label{prod}
\eeq
and these sterile neutrinos come in thermal equilibrium only if
$\theta > 5 \times 10^{-4} \left(1~\mbox{keV}/M_s\right)^{1/2}$.

Another limit comes from the observation \cite{Dolgov:2000ew} that
radiative decays $N_1 \to \gamma\nu,\gamma\bar{\nu}$ can be observed
as a feature in the spectrum of cosmic X-ray radiation. This effect
is similar to the ultra-violet radiation resulting from decays of
active neutrinos, proposed in  \cite{DeRujula:1980qd} as a possible
signal from dark matter composed of ordinary neutrinos. According to
Ref.~\cite{Abazajian:2001vt,Abazajian:2005gj}, the analysis of X-ray
emission from  Virgo cluster gives a constraint $\theta < 1.6\times
10^{-3}\left(1~\mbox{keV}/M_s\right)^2$, for $1~\mbox{keV}< M_s <
10~\mbox{keV}$, whereas the study of the diffuse X-ray background
\cite{Boyarsky:2005us} gives $\theta < 5.8 \times
10^{-3}\left(1~\mbox{keV}/ M_s \right)^{5/2}$, for  $1~\mbox{keV} <
M_s < 100~ \mbox{keV}$.

Further constraints on the parameters of dark-matter sterile neutrino
depend on the assumptions about the initial concentration of $N_1$ at
some high energy scale which we will take to be much larger than $M_W
\sim 100$ GeV. We are going to distinguish three different scenarios.

I. Sterile neutrinos are not produced at high temperatures, i.e. 
their abundance is zero at $T > 1$ GeV. This can happen if the
couplings  of the singlet fermions to the fields beyond the Standard
Model, including the inflaton, are sufficiently weak. In this case it
is required that $\theta = \theta_{max}(M_s)$ in order to produce the
right amount of dark matter 
\cite{Dodelson:1993je,Dolgov:2000ew,Abazajian:2001nj,Abazajian:2005gj}.
This requirement, together with an upper bound on the angle $\theta$
from the Virgo cluster observations~\cite{Abazajian:2001vt}, gives an
upper limit on the mass, which is, according to
\cite{Abazajian:2005gj},  $M_s  <  8$~keV.

The sterile neutrinos produced in neutrino oscillations have a
non-negligible free-streaming length.  For masses below
$M_{min}=2$~keV, the free-streaming would erase small-scale structure
known to exist from cosmic microwave background radiation,
Lyman-$\alpha$ forest, and Sloan Digital Sky Survey
data~\cite{Hansen:2001zv,Viel:2005qj,Abazajian:2005xn}. Therefore,
the allowed range of masses is $2\, {\rm keV} < M_s < 8\, {\rm
keV}$. 

II. Sterile neutrinos were in thermal equilibrium at high 
temperature, and their abundance at the electroweak scale coincides
with that of the active neutrinos.  This case is excluded since, even
if the constraint (\ref{angle}) is satisfied, the present
concentration of sterile neutrinos is only a factor
$g^*(M_W)/g^*(T_d) \sim {\cal O}(10)$ smaller than the concentration
of active neutrinos, where $g^*(T)$  is the number of effectively
massless degrees of freedom at temperature $T$, $T_d \simeq 1$ MeV is
the decoupling temperature of active neutrinos. The mass of the
sterile neutrino, required in this case, $M_s \sim 100$ eV, is at
odds with the structure formation and  with a very conservative 
Tremaine-Gunn bound \cite{Tremaine:1979we} on the mass of any
fermionic dark matter, which requires $M_s > M_{TG} \simeq
0.3~\mbox{keV}$ when applied to the dwarf spheroidal galaxies
\cite{Lin:1983vq,Dalcanton:2000hn}. 

III.  The sterile neutrinos are produced at some high energy scale in
just  the right amount to be dark matter by some processes that have
nothing to do with the active-sterile neutrino oscillations. This
scenario, though quite trivial,  has not been considered previously.
In this case, there is no upper limit on their mass, while the lower
limit is as small as $M_{TG}$.  The angle $\theta$ is bounded from
above by eq.~(\ref{angle}) and by the X-ray observations\footnote{For
$\theta < 3\times 10^{-6}$ the sterile neutrinos cannot explain the
pulsar kicks and the early reionization.}.

The mixing angles $\theta<2\times 10^{-4}$, required for a dark
matter sterile neutrino in this model, leave little hope for a direct
detection of sterile neutrinos in a laboratory experiment
\footnote{Of course, none of these bounds apply if the universe has
undergone a low scale inflation and was never reheated to a
temperature above MeV~\cite{Gelmini:2004ah}. Even in this case, one
could envision some way to produce dark matter in the form of sterile
neutrinos, for example, from a direct coupling of the inflaton to
$N_1$. }.  The most promising detection strategy is based on the
observation of X-rays from sterile neutrino
decays~\cite{Dolgov:2000ew,Abazajian:2001vt,Boyarsky:2005us,Abazajian:2005gj}. 

Let us now discuss how these constraints are modified in the
$\nu$MSM.

{\bf The model.} The Lagrangian of the  $\nu$MSM contains the
Standard Model fields and gauge singlets $N_I$, as well as the
following set of terms: 
\begin{eqnarray}
  {\cal L}_{\nu \rm{MSM}} & = & {\cal L}_{\rm MSM}
  + \ol N_I i \partial_\mu \gamma^\mu N_I \nonumber \\ 
& &   - F_{\alpha I} \, \Phi  \ol L_\alpha N_I 
  - \frac{M_I}{2} \; \ol {N_I^c} N_I + \rm{h.c.}  \, ,
  \label{lagr}
\end{eqnarray}
where $\Phi$ and $L_\alpha$ ($\alpha=e,\mu,\tau$) are the Higgs and
lepton doublets, respectively, $\langle \Phi \rangle=v=174$ GeV is
the Higgs vacuum expectation value, $M^D = F v$ and $M_I$ are Dirac
and Majorana masses, respectively. An obvious requirement for the
parameters of the  $\nu$MSM is that the masses of active neutrinos,
given by the see-saw expression
\beq
m_\nu = - M^D\frac{1}{M_I}[M^D]^T, 
\label{eq:Mseesaw}
\eeq
must be in agreement with neutrino oscillation experiments. We will
assume that this model is a viable effective field theory up to some
very high energy scale, which can be as large as the Planck scale,
but is not smaller than ${\cal O}(1)$ TeV or so, which is required
for the validity of the baryogenesis mechanism via sterile neutrino
oscillations. The mixing angle $\theta$ is expressed through the
parameters of the $\nu$MSM as
\beq
\theta^2 = \frac{1}{ M_s^2} \sum_{\alpha =e\mu\tau}|M^D_{\alpha
1}|^2~,
\eeq
where $M_s$ is the lightest Majorana mass $M_1$. The masses of the
heavier sterile neutrinos are free parameters of the model. To
explain the baryon asymmetry of the universe, $N_2$ and $N_3$  must
be highly degenerate in mass to amplify the coherent CP-violating
effects in sterile neutrino oscillations. Moreover, their (common)
mass $M$ cannot exceed greatly the electroweak scale since large
Majorana masses enhance the rate of the lepton number
non-conservation, which leads to a dilution of the baryon asymmetry. 

{\bf Constraints on the sterile neutrinos in the $\nu$MSM.} For 
reasons explained below we will take $M \sim {\cal O}(1-10)$ GeV. 
There are limits on sterile neutrinos for masses below
400~MeV~\cite{bounds_on_sterile_nu}, but for larger masses there are
no laboratory constraints.  

As in the case of the Standard Model with one extra singlet fermion,
to discuss the allowed range of parameters, one has to fix the
concentrations of extra sterile neutrinos at temperatures higher than
the electroweak scale. The analysis of all possible cases is beyond
the scope of this paper. We will assume here that at least one of 
the heavy states $N_{2,3}$ was in thermal equilibrium at some
temperature greater than $M$. This assumption is certainly true if
the Yukawa coupling constants $f_I^2=\sum_{\alpha}|F_{\alpha I}|^2$
are large enough. Heavier sterile neutrinos  can be created in the
Higgs decays or in collisions of $t$-quarks at the electroweak scale,
and they  equilibrate at $T\sim M_W$ if  $f_I^2 > 2\times 10^{-14}$
\cite{Akhmedov:1998qx}. For smaller Yukawa couplings 
\beq
f_I^2 >
10^{-19}\left(\frac{M}{1~\mbox{GeV}}\right),
\label{small}
\eeq
they equilibrate at $T_D \sim {\cal O}(20)$ GeV, according to
eq.~(\ref{prod}) (the estimate (\ref{small}) is valid for  masses
$1~\rm{GeV} < M < 10~\rm{GeV}$). The Yukawa couplings even smaller
than these are allowed if $N_2$ and $N_3$ have sufficiently strong
additional interactions at some high energy scale.

The concentration of $N_2$ and $N_3$ at time of their decay,
corresponding to temperatures $T_r\simeq {\cal O}(1)$ MeV (see below) 
is given by 
\beq
n_{2,3}=\frac{g^*(T_r)}{g^*(T_D)}\frac{6\zeta(3)g_N^*}{7\pi^2} T^3~,
\eeq
where $g_N^* = 2\times \frac{7}{8}$ is the degeneracy factor for
$N_{2,3}$, and we take  $g^*(T_r) =10.75$ and $g^*(T_D)=86.25$, which
accounts for quark degrees of freedom up to the $b$ quark. This
equation is valid, provided eq.~(\ref{small}) is satisfied. For
smaller Yukawa couplings, according to our assumption, one should
replace $g^*(T_D)$ by $g^*(M_W) \simeq 106.75$.

The sterile neutrinos $N_{2,3}$ are unstable and can decay in many
channels, depending on their mass. In general, $N_{2,3}$ decays lead
to the entropy production, given by a  factor  $S > 1$. If the decay
temperature is {\em smaller} than the temperature at which the
dark-matter  sterile neutrino $N_1$ is produced, the density of $N_1$
is diluted by $S$ and also the  momentum distribution is redshifted
by a factor $S^{\frac{1}{3}}$.  This dilution allows for a wider
range of allowed masses, with the lower bound $M_s >
M_{min}/S^{\frac{1}{3}}$. In addition to this, the constraint on the
mixing angle $\theta^2$ becomes weaker by a factor $S$. Moreover, at
sufficiently large $S$ even an equilibrium concentration of sterile
neutrino $N_1$ can be sufficiently diluted. This means that the
mixing angle is limited only by the X-ray constraints 
\cite{Abazajian:2001vt,Abazajian:2005gj,Boyarsky:2005us} discussed
above. 

Let us show that the entropy production factor as large as $S\sim
{\cal O}(100)$ does not affect the predictions of the  Big Bang
nucleosynthesis (BBN).

The ratio of entropy produced from the $N_{2,3}$ decays to the rest
of the entropy is  $S=[g^*(RT)^3]_{\rm final}/[g^*(RT)^3]_{\rm
initial}$, where $R$ is the scale factor. The value of  $S$ depends
on the dimensionless combination $A=\Gamma M_0 (g^*(T_D)/M)^2$, were
$\Gamma$ is the width of the decaying particle, and  $M_0=2.4\times
10^{18}$ GeV is the reduced Planck mass. It was computed in
\cite{Scherrer:1984fd} for different situations. For our case it is
given by 
\begin{equation}
S \simeq \left[1+(1.37/A)^{2/3}\right]^{3/4} ~.
\end{equation}

To find what maximal allowed value of $S$, the Big Bang
nucleosynthesis should be reanalyzed in the presence of $N_{2,3}$
decays, taking into account the specific branching ratios for
different final states.  A similar problem has been addressed in a
number of
papers~\cite{Kawasaki:1999na,Kawasaki:2000en,Hannestad:2004px,
Ichikawa:2005vw}. In these papers the initial state of the universe
before the BBN was assumed to contain  nothing but a decaying scalar
field, which produced  particles of the Standard Model and created
all the entropy. In other words, the factor $S$  was assumed to be
very large. In our case $N_{2,3}$ decay in plasma, and, therefore,
the constraints must be weaker.  We will  take a conservative
approach and adopt the constraints from
Refs.~\cite{Kawasaki:1999na,Kawasaki:2000en,Hannestad:2004px,Ichikawa:2005vw},
even though they are too strong for our case. 

According to \cite{Kawasaki:2000en}, the reheating temperature,
defined as $T_r = 0.554 \sqrt{\Gamma M_0}$, can be as small as $0.7$
MeV without a conflict with BBN. In this case the effective number of
active neutrino degrees of freedom at the time of BBN is less than
three.  There is some indication that this may, indeed, be required
for the BBN predictions to agree with
observations~\cite{Steigman:2005uz}. A more stringent constraint $T_r
> 4$ MeV was derived in Ref.~\cite{Hannestad:2004px}, where the
information about CMB anisotropies and matter distributions was
incorporated into the analysis.  We will present our results for both
$T_r = 0.7$ MeV and $T_r = 4$ MeV.

Since the dilution of dark matter depends on the parameter $A$, let
us determine the allowed range of decay widths $\Gamma_{2}$ and
$\Gamma_{3}$ of the heavy sterile neutrinos $N_2$ and $N_3$,
respectively.   The neutral current diagrams ($Z$ exchange) give the
decay modes $N_{2,3} \to \nu \nu \bar \nu,~ \nu lPlot[((1/y[x]) /. sol), {x, xin, xout}] \bar l,~  \nu q
\bar q$ plus charge conjugated channels, while the charged currents
($W$ exchange) channels produce $l q \bar q$ in the final state (here
$q$ corresponds to quarks and $l$ to charged leptons).  In general,
these decay widths are given by
\bea
\Gamma_I =\frac{G_F^2 M^3}{192 \pi^3}
\sum_{\alpha} A_{\alpha}|M^D_{\alpha I}|^2~,
      \label{width}
\eea
where $G_F$ is the Fermi constant, and the coefficients $A_\alpha
\sim 1$ depend on the kinematically allowed channels. For example,
for a $10$ GeV sterile neutrino one can show that 
%\beq
$\Gamma_I < 22 G_F^2 M^3 f_I^2 v^2/(192 \pi^3)~.$
%\eeq

The smaller is the Yukawa coupling $f_I$, the more entropy production
one can expect.   However, not all choices of the Yukawa couplings
are consistent with the data on neutrino oscillations, in particular
with the  mass square differences. Applying the general analysis of
the see-saw formula made in \cite{Asaka:2005an}, one can prove that
it is impossible to have very small $f_{2,3}^2 \ll f_{sol}^2$ and to
reproduce the data (here $f_{sol}^2= \sqrt{\Delta m^2_{sol}} M/v^2
\simeq 3\times 10^{-16}M/\mbox{GeV}~, \sqrt{\Delta m^2_{sol}}\simeq 9
\times 10^{-3}$ eV). In other words, only one of the particles
$N_{2,3}$ can reheat the universe considerably. We take it to be
$N_{3}$ for definiteness and choose  $f_{3}^2 \ll f_{sol}^2$. There
is no lower bound on $f_{3}^2$ from experiment. However, in order to
reproduce the observed mass differences one must have $f_1^2 v^2/M_s$
and $f_2^2 v^2/M$ of the order of $\sqrt{\Delta m^2_{sol,atm}}\sim
{\cal O}(10^{-2})$~eV. Moreover, repeating the arguments of
Ref.~\cite{Asaka:2005an}, one can prove that for such a small value
of $f_3$ the lightest active neutrino must be much lighter  than
$\sqrt{\Delta m^2_{sol}}$. Hence, the predictions of the active
neutrino masses made in Ref.~\cite{Asaka:2005an} remain
valid\footnote{The analysis of the active neutrino masses made in
Ref.~\cite{Asaka:2005an}  did not take into account the possibility
of reheating of the universe due to heavy sterile neutrino decays.}. 

Now we are ready to proceed to numerical estimates of the entropy
generation factor $S$. Taking the Yukawa coupling from
eq.~(\ref{small}), we get that for $M=5~(11)$ GeV and $S=29~(10)$ for
the reheating temperature $0.7~(4)$ MeV. In this case the heavy
sterile neutrino decoupling temperature is ${\cal O}(20)$ GeV.  At
temperature  ${\cal O}(1)$~MeV, the energy density of sterile
neutrinos dominates by a factor $10$ or $4$, respectively.   For
smaller Yukawa couplings, $N_3$ is out of thermal equilibrium for all
temperatures in the $\nu$MSM. Of course, the $N_3$ abundance can be
large due to some physics beyond the $\nu$MSM. Then, for $f_3 <
10^{-10}$ the entropy production factor can exceed $S=100$, with the
energy dominance factor greater than $20$.

This range of $S$ opens new possibilities for the sterile neutrino
production. First, $N_1$ can be produced from neutrino oscillations,
as in
Refs.~\cite{Dodelson:1993je,Dolgov:2000ew,Abazajian:2001nj,Abazajian:2005gj},
but for some larger mixing angles $\theta$. As long as the lightest
neutrinos are never in equilibrium in the early universe, the amount
of dark matter produced from neutrino oscillations is proportional to
$\theta^2$ \cite{Dodelson:1993je}. Now, the admissible angle can be
larger than (\ref{angle}) by a factor of $\sqrt{S}$.

The second possibility is an even more drastic departure from the
usual scenario. The lightest sterile neutrinos could, in fact,
achieve the  equilibrium density in the early universe due to some 
high-scale physics, new interactions, or because the mixing angle
$\theta$  is large enough. Then at some temperature  $T_{\rm d,s}$
sterile neutrinos freeze out from equilibrium.  Their density would
then be 
\beq
\Omega_s = 0.26 \left ( \frac{M_s}{12\, {\rm eV}} \right ) \frac{1}{S}
 \left (\frac{g^*(T_d)}{g^*(T_{\rm d,s})} \right ). 
\eeq
This gives the right dark matter density for $M_s \sim
\frac{g^*(T_{\rm d,s})}{g^*(T_d)} \frac{S}{100}\, {\rm keV}$.

In the absence of  entropy production, masses below 2~keV correspond
to dark matter that is too warm.  However, the dilution and red shift
also change the free-streaming length of the sterile neutrinos by a
factor $S^{1/3}$.  Let us, therefore, consider the limits from
large-scale structure in more detail. 

Some bounds on warm dark matter come from the observation of
small-scale structure in the Lyman-$\alpha$ forest, as well as from
SDSS data.  The small-scale structure is sensitive to the
free-streaming length of sterile neutrinos, $\lambda_{\rm FS} =
\int_0^t (v(t')/R(t')) dt'$, where $v$ is the velocity of sterile
neutrinos.  For the parameters of interest, the lightest sterile
neutrino becomes non-relativistic at the time $t_{_{\rm NR}}$
(redshift $z_{_{\rm NR}}$) long after the decay of the heavier
sterile neutrinos and before the matter-radiation equality, which
occurs at $t_{\rm eq}$. Therefore, one can neglect the contribution
to $\lambda_{\rm FS}$ from the matter-dominated phase prior to heavy
neutrino decay, and the integral for the free-streaming length has
the usual expression:
\begin{eqnarray}
\lambda_{\rm FS} &\approx & t_{_{\rm NR}} z_{_{\rm NR}} [2+\ln(t_{\rm
eq}/t_{_{\rm NR}})] \nonumber \\
& \sim & 2 \, {\rm Mpc} \left ( \frac{\rm keV}{M_s}\right ) 
\left ( \frac{T_s}{T_\nu}  \right ) \, 
\left ( \frac{1}{S}  \right )^{1/3}, 
\end{eqnarray}
where $T_s$ is the temperature of the dark-matter sterile neutrinos
in the absence of dilution and $T_\nu$ is the temperature of active
neutrinos. 

For a numerical example let us assume that the Lyman-$\alpha$
constraint in the absence of dilution is $M_s>M_{\rm min}\simeq 2$
keV \cite{Viel:2005qj}. For Scenario I we have $T_s \simeq T_\nu$,
and for $S \simeq 13$ all masses $M_s>0.8$ keV are allowed. In this
case the sterile neutrinos are out of thermal equilibrium and their
mixing angle is smaller than $\theta < 5.4\times 10^{-4}$.

For larger mixing angles sterile neutrinos come to thermal
equilibrium and can be considered as the ``generic warm dark matter".

The maximal possible mixing angle in this case comes from
consideration of active neutrino masses and is of order of
$\theta^2 \sim \sqrt{ \Delta m_{atm}^2 }/M_{\rm min}
\simeq 8.5 \times 10^{-5}$ where $M_{\rm min} = 0.55$ keV 
so that the free streaming length becomes too small to affect 
the observed structure \cite{Viel:2005qj}.
Then, in this case, sterile neutrino can be the dark matter
for $( g^\ast (T_{d,s})/g^\ast (T_d) ) S \simeq 55$.

We, therefore, conclude that, for large enough $S$, sterile neutrinos
can be the dark matter for masses $M_s > 0.55$~keV and mixing angles
$\theta > \theta_{max}(M_s)$.  These mixing angles are probably
within the reach of laboratory experiments \cite{Gelmini:2004ah}.

The late entropy production opens a possibility to have several
component WDM even with a single dark matter candidate. Namely, a
fraction of sterile neutrinos could be created above the electroweak
scale, and then a separate population of them could be produced in
active--sterile neutrino oscillations. The two populations can have
effective temperatures  different by a factor of $S^{\frac{1}{3}}$,
leaving a fingerprint on a distribution of matter at scales of the
order of a  few Mpc.  

Some comments are now in order.  (i) The dilution of the sterile
neutrino abundance leads to the decrease of the baryon asymmetry at
the same time. This effect, however, can be compensated by  an
increase in the degeneracy of the states $N_2$ and $N_3$.  (ii) The
mixing angle $\theta$ is still bounded from above by the X-ray
observations \cite{Abazajian:2001vt,Boyarsky:2005us}. These limits,
however, are very weak for $M_s < 1$ keV, as discussed in  
Ref.~\cite{Gelmini:2004ah}. (iii) The large value of the total
entropy production factor in the $\nu$MSM opens a possibility to have
any type of a warm dark matter candidate with mass being larger than
$0.55$ keV with decoupling temperature above the electroweak scale,
including the light gravitino or other weakly interacting massive
particles~\cite{McDonald:1989jd}. (iv) The scenario III, in which the
dark matter sterile neutrinos are produced not in the active--sterile
transitions but in some other processes, is valid for the $\nu$MSM as
well. In this case the mixing angle $\theta$ is constrained by the
X-ray observations, while the mass $M_s$ should satisfy the
Tremaine-Gunn bound or the Lyman-$\alpha$ bound, depending on the
mechanism of their production. 

{\bf Conclusions.} To summarize, in the Standard Model extended by
three singlet fermions ($\nu$MSM), the lightest sterile neutrino is a
viable dark-matter candidate. For the case when the initial
concentration of sterile neutrinos is assumed to be zero, the main
reason allowing for a new window for WDM parameters is the decay of a
heavier neutrino before the BBN epoch. These decays can produce
enough entropy to (i) dilute the density of sterile neutrinos and
(ii) reduce their temperature.  The second effect, cooling of the
sterile neutrino spectrum, reduces the free-streaming length and
weakens the bounds from structure formation on non-thermal sterile
neutrino.  In addition, a sterile neutrino with any mass satisfying
the Lyman-$\alpha$ bound, that was produced by a mechanism not
related to active--sterile oscillations, is perfectly allowed.

As has been already discussed in
\cite{Asaka:2005an,Asaka:2005pn,Boyarsky:2006jm}, the $\nu$MSM with
three singlet fermions can be falsified by the data in neutrino
physics. In-spite of the efforts of two authors of the present paper
(T.A. and M.S.) a range of parameters in which the $\nu$MSM can give
the dark matter, explain the baryon asymmetry of the Universe, and
the data on neutrino oscillations {\it including} the LSND anomaly,
has not been found. The LSND anomaly can be accommodated to the
$\nu$MSM, provided the dark matter or the baryon asymmetry
explanations are given up. Thus, if MiniBoone experiment confirms the
LSND results, the $\nu$MSM should be extended by at least one extra
sterile neutrino, if the requirements to be consistent with cosmology
are kept. The same conclusion is true if the active neutrinos are
found to be degenerate in mass.

The work of A.K. was supported in part by the DOE grant
DE-FG03-91ER40662 and by the NASA ATP grants NAG~5-10842 and
NAG~5-13399. The work of T.A. was supported by the Grant-in-Aid for
Scientific Research from the Ministry of Education, Science, Sports,
and Culture of Japan, No. 16081202. The work of M.S. was supported in
part by the Swiss Science Foundation.  M.S. thanks  Alexey Boyarsky
and  Oleg Ruchayskiy for many helpful discussions and to the UCLA,
where part of this work was done, for hospitality. We thank Oleg
Ruchayskiy for kindly providing us a figure with constraints on the
properties of sterile neutrino.

{\bf Note added.} After this work has been submitted for publication,
a number of papers on sterile neutrinos appeared
\cite{Seljak:2006qw,Boyarsky:2006zi,Boyarsky:2006fg,Riemer-Sorensen:2006fh,Shaposhnikov:2006xi}.
In \cite{Seljak:2006qw} a new analysis of the Lyman-$\alpha$ forest
data has been performed. In
\cite{Boyarsky:2006zi,Boyarsky:2006fg,Riemer-Sorensen:2006fh} new
constraints on the mixing angle $\theta$, coming from the analysis of
X-rays from galaxy clusters and from our own galaxy have been
derived. These results represent a considerable improvement of the
constraints existing previously and discussed in the text.

\begin{figure}[t]
  \centering %
  \includegraphics[width=\linewidth]{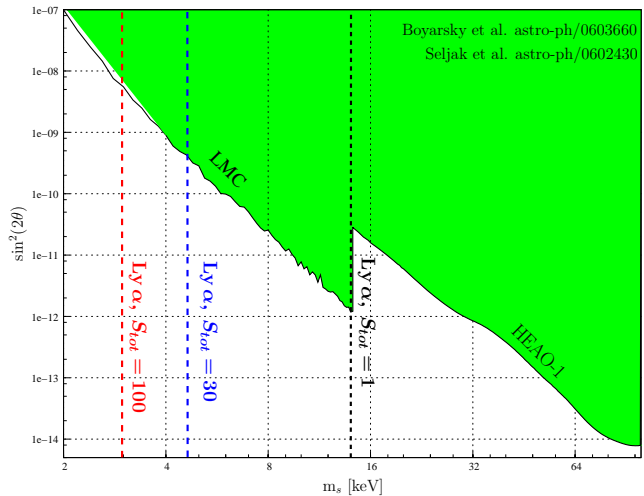} % 
  \caption{Restrictions on the mixing angle of sterile neutrino
  coming from X-ray observations of \cite{Boyarsky:2006fg} and from
  the Lyman-$\alpha$ forest data analysis of \cite{Seljak:2006qw},
  assuming that all dark matter is in sterile neutrinos.  No
  region is allowed for sterile neutrinos produced in active-sterile
  transitions, if their initial concentration was zero.  For $N_1$
  produced thermally at high energy scale the allowed region is for
  $M_s > 2.5$ keV and below the shaded area. For primordial
  production of sterile neutrinos in light inflaton decays
  \cite{Shaposhnikov:2006xi}, all values below shaded area and to the
  right from $M_s =10$ keV are allowed.}
  \label{fig1}
\end{figure}

In Fig. \ref{fig1} we show the limits from  \cite{Boyarsky:2006fg},
based on the statistical analysis of the X-ray data  with the method
described in \cite{Boyarsky:2005us}, and the constraints from
\cite{Seljak:2006qw}. The limits on the mixing angle, derived in
\cite{Riemer-Sorensen:2006fh}, are coming from the requirement that
the flux in the dark matter emission line is smaller than the total
observed flux. These constraints are weaker than those of ref. 
\cite{Boyarsky:2006fg} and allow more space for sterile neutrinos, as
discussed in  \cite{Riemer-Sorensen:2006fh}.
%%%%%%%%%%%%%%%%%%%%%%%%%%%%%%%%%%%%%%%
%%%%%%%%%%%%%%%%%%%%%%%%%%%%%%%%%%%%%%%

A combination of new X-ray and Lyman-$\alpha$ bounds can rule out a
number of scenarios discussed in this paper, unless eq.
(\ref{angle}), derived in \cite{Abazajian:2005gj}, gets strongly
modified by largely unknown dynamics of the strongly interacting
plasma at temperatures $T\sim 150$ MeV or the bound of
\cite{Seljak:2006qw}, $M_s>m_0/S_{\rm tot}^{1/3}$ with $m_0 \simeq
14$ keV, involving complicated hydrodynamical simulations, gets
relaxed because of some reason. Here factor $S_{tot}$ is defined as
$S_{\rm tot}=(g^*(T_{d,s})/g^*(T_d))S$

If $m_0$ is as large as in \cite{Seljak:2006qw} ($m_0 \simeq 2$ keV
according to the earlier works, as discussed in the text), the
Dodelson-Widrow scenario (scenario I of this work), is excluded. 
Indeed, if eq. (\ref{angle}) is combined with the X-ray data, the
upper limit of $8$ keV, discussed in the text, is replaced by $2.8$
keV \cite{Boyarsky:2006fg}, which is well below the Lyman-$\alpha$
bound. The scenario II, which was impossible for the MSM with
addition of one sterile neutrino and possible for the $\nu$MSM, would
require to have the entropy production factor greater than $S \simeq
25$ and sterile neutrino mass greater than  $m_s \simeq 2.5$ keV.  On
the contrary, if the concentration of the dark matter sterile
neutrino is postulated to be zero at temperatures above few hundreds
MeV, then even the $\nu$MSM with entropy production cannot reconcile
the new constraints. In particular, the scenario with large mixing
angle is excluded (it would be allowed if $m_0 \simeq 5$ keV). If
true, this points out towards the production mechanism of sterile
neutrinos, not related to active-sterile neutrino oscillations
(scenario III of the present work). Indeed, in
\cite{Shaposhnikov:2006xi} was demonstrated that the inflaton decays
can produce effectively the dark matter sterile neutrinos which
satisfies all the constraints of 
\cite{Seljak:2006qw,Boyarsky:2006zi,Boyarsky:2006fg,Riemer-Sorensen:2006fh}.

\end{document}